\documentclass[runningheads]{llncs}
\usepackage[T1]{fontenc}
\usepackage{amsmath}
\usepackage{amsfonts}
\usepackage{tikz}
\usetikzlibrary{arrows.meta, positioning, calc}
\usepackage{tabularx}
\usepackage{caption}
\usepackage{makecell}
\usepackage{multirow}
\usepackage{dsfont}
\usepackage{subcaption}
\usepackage{graphicx}
\usepackage{float}
\usepackage{booktabs}
\usepackage{environ}
\usepackage{float}
\usepackage{amsmath}
\usepackage{comment}
\usepackage{amssymb}
\usepackage{diagbox}
\usepackage[numbers]{natbib}
\usepackage{bm}
\usepackage{environ}
\usepackage{multicol}
\usepackage{xcolor}
\usepackage{mathtools}
\usepackage{amssymb}
\usepackage{enumitem}  

\newcommand{\N}{\mathbb{N}}

\newcommand{\mAi}{\mathcal{A}^i}

\allowdisplaybreaks[3]

\newcommand{\ha}{\hat{a}}
\newcommand{\mc}{\mathcal}

\allowdisplaybreaks[3]

\usepackage[font=small,labelfont=bf,labelsep=space]{caption}
\usepackage[hypertexnames=true]{hyperref}
\usepackage{graphicx}
\begin{document}
\title{On the characterization of constrained correlated equilibria in Markov games} 
%
%
\author{Tingting Ni\inst{1,2} \and
Anna Maddux\inst{1,2} \and
Maryam Kamgarpour\inst{1}}

\authorrunning{T. Ni et al.}

\institute{SYCAMORE, EPFL \and
Equal contribution}
\maketitle              
\begin{abstract}
Markov games with coupling constraints model constrained dynamical decision-making involving self-interested agents, where the feasibility of an individual agent’s strategy depends on the joint strategies of the others. Such games arise in numerous real-world applications involving safety requirements and budget caps, for example, in environmental management, electricity markets, and transportation systems. In unconstrained dynamical decision-making, the correlated equilibrium has emerged as a desired solution concept, due to its computational tractability and amenability to learning algorithms. Understanding how coupling constraints shape correlated equilibria is a crucial step towards computing solutions in constrained Markov games. In this paper, we formalize and characterize the notion of \emph{constrained correlated equilibria} for Markov games, defined as feasible joint policies where any unilateral deviation is either unprofitable or infeasible. 

\keywords{Markov games   \and Correlated equilibria \and Coupling constraints}
\end{abstract}

\section{Introduction}

Many real-world systems involve multiple self-interested agents that interact with each other in a dynamic environment. Such multi-agent systems can be framed as Markov games, also referred to as stochastic games \cite{shapley1953stochastic}, which extend normal-form games to dynamic settings. With the rapid rise of multi-agent reinforcement learning, Markov games have become a central framework for studying coordination, competition, and learning among agents. While multi-agent reinforcement learning has achieved impressive success in handling complex and high-dimensional systems \cite{zhang2021multi,schrittwieser2020mastering}, safety concerns arise when applying it to real-world systems, where agents often face coupling constraints—that is, the feasibility of each agent’s strategy depends on the joint actions of all others. 

Prominent examples arise in environmental management~\cite{madani2017serious}, where countries must jointly ensure that greenhouse emissions are below some threshold, in electricity markets~\cite{visudhiphan1999dynamic}, where transmission capacity constraints must be satisfied, or in transportation systems~\cite{mylvaganam2017differential}, where vehicles must avoid collisions with other vehicles. Such multi-agent systems can be framed as Markov games with coupling constraints \cite{debreu1952social}, representing a direct generalization of constrained Markov decision processes to the multi-agent setting~\cite{altman2021constrained, chen2004dynamic, feinberg2020constrained}. To analyze constrained multi-agent decision making, it is crucial to characterize constrained equilibrium notions for this class of games and prove its existence. 

%
%
%

The \emph{generalized Nash equilibrium} was first introduced in~\cite{rosen1965existence} for normal-games as a natural extension of its unconstrained counterpart, the Nash equilibrium \cite{nash1951non}. Several works have since emerged to study the generalized Nash equilibrium, including its characterization and existence in various game settings, ranging from constrained normal-form games~\cite{rosen1965existence,facchinei2007generalized, balbus2008existence,dutang2013existence,kulkarni2012variational,yin2011nash,fischer2014generalized, braouezec2023economic,tian1992existence} to constrained Markov games \cite{altman2000constrained, alvarez2006existence,dufour2022stationary,dufour2024nash,zhang2018continuous}. However, without further assumptions, computing a generalized Nash equilibrium is intractable \cite{daskalakis2009complexity}, a property it inherits from the Nash equilibrium.  

To circumvent intractability of Nash equilibria in normal form games, past works have turned to weaker equilibrium notions. In particular, a correlated equilibrium \cite{aumann1974subjectivity, aumann1987correlated} is a strict generalization of Nash equilibria which can be computed and learned efficiently \cite{li2024coupled,misra2023robust,cesa2006prediction,bubeck2012regret,zhang2025learning}. In Markov games, tractability of correlated equilibria has also recently been established \cite{daskalakis2023complexity}. However, to date, equilibrium notions in \emph{constrained} Markov games have received little attention. This motivates studying the constrained correlated equilibrium which was recently introduced by~\cite{chen2022finding,boufous2024constrained,bernasconi2023constrained,jaskiewicz2023approximate}.

In constrained Markov games, a constrained correlated equilibrium is a feasible policy such that any \emph{unilateral modification} is either unprofitable or leads to an infeasible outcome. Different classes of modifications—deterministic versus stochastic—give rise to alternative equilibrium formulations. In particular, stochastic modifications form a more general class of modifications, and therefore define a stronger equilibrium concept. Characterizing constrained correlated equilibria under these different classes of modifications remains an open research problem.  In the unconstrained setting, both for normal-form and Markov games, it was shown that restricting to the subset of deterministic modifications yields an equivalent formulation of a correlated equilibrium. However, in the presence of constraints, such a restriction leads to a weaker equilibrium notion \cite{chen2022finding}. As the structure and characterization of constrained correlated equilibria in terms of possible modifications remain only partially understood, we are led to the following question:
\begin{center}
    \textit{Which classes of modification yield equivalent notions of constrained correlated equilibria?}
\end{center}

Our paper addresses the characterization of constrained correlated equilibria in finite-horizon Markov games with finite state and action spaces. For clarity, we summarize our results alongside existing results on characterization in both normal-form and Markov games in Table~\ref{table}. Our main contributions are as follows:
\begin{enumerate}
    \item  We show that constrained correlated equilibria are equivalently characterized by restricting to convex combinations of deterministic modifications (Theorem~\ref{cor:equinotion}). 
\end{enumerate}
\begin{table}[t]
\centering
\small
\renewcommand{\arraystretch}{1.25}
\setlength{\tabcolsep}{4pt}
\begin{tabularx}{\textwidth}{@{}p{0.22\textwidth}XX@{}}
\toprule
\textbf{Equilibrium} 
& \textbf{Normal-form games} 
& \textbf{Markov games} \\
\midrule

Correlated equilibrium
& Deterministic modifications~\cite{aumann1987correlated}
& Deterministic modifications~\cite{chen2022finding} \\

\midrule

Constrained correlated equilibrium
& \multicolumn{2}{p{0.70\textwidth}@{}}{
Stochastic modifications~\cite{chen2022finding}; 
convex combinations of deterministic modifications (see \textbf{Theorem~\ref{cor:equinotion}}).
} \\

\bottomrule
\end{tabularx}
\caption{Characterization of (constrained) correlated equilibria in normal-form and Markov games}
\label{table}
\end{table}


\noindent \textbf{Notation} Let $\mathbb{N}$ and $\mathbb{R}$ denote the sets of natural and real numbers, respectively. For any $m \in \mathbb{N}$, we define $[m] := \{1, \dots, m\}$. We denote by $x_{m:n}$ the sequence $\{x_m, x_{m+1}, \dots, x_n\}$. Let $\mathbf{1}_n$ denote the all-ones vector of dimension $n$. Given a finite set $\mathcal{X}$, we denote the indicator function by $\mathds{1}_{\mc X}(x)$. Furthermore, the probability simplex over $\mc X$ is denoted  by \(\Delta({\mathcal{X}})\) and the cardinality of $\mc X$ is denoted by \(|\mathcal{X}|\). 
\section{A finite-horizon Markov game with coupling constraints} \label{sec:Markov_game}

A finite-horizon Markov game with coupling constraints is given by the tuple \( \{\mc N, H, \mc S, \{\mc A^i\}_{i\in\mc N},  P, \rho,\{r^i\}_{i\in\mc N}, \{g^{i,j}\}_{i\in\mc N,j\in[J]}, \{c^{i,j}\}_{i\in\mc N,j\in[J]}\}\), where $\mc N=[N]$ denotes the set of players and $H\in\N$ denotes a finite horizon. The set \(\mc S\) denotes a finite state space and each player $i\in\mc N$ has a finite action space $\mc A^i$. Let $a=(a^1,\ldots, a^N)\in\mc A$ denote the joint action profile, where $\mc A =\Pi_{i=1}^N A^i$ is the joint action space. Similarly, let $a^{-i}=(a^1,\ldots,a^{i-1},a^{i+1},\ldots,a^N)\in\mc A^{-i}=\Pi_{j\neq i}\mc A^j$ denote the joint action profile of all players except player $i$. The transition kernel is denoted by \( P := \{P_t\}_{t\in[H-1]} \), where \( P_t : \mathcal{S} \times \mathcal{A} \rightarrow \Delta(\mathcal{S}) \). Specifically, \( P_t(s_{t+1} |s_t, a_t) \) denotes the probability of transitioning from state \( s_t \in \mathcal{S}\) to state \( s_{t+1} \in \mathcal{S} \) under the joint action \( a_t \in \mathcal{A} \). The initial state distribution is denoted by $\rho\in\Delta(\mc S)$.

Furthermore, each player \( i \in \mathcal{N} \) has a reward function \( r^i := \{r^i_t\}_{t \in [H]} \), where \( r^i_t : \mathcal{S} \times \mathcal{A} \rightarrow [0,1] \). Additionally, each player has \( J \) constraint functions \( g^{i,j} := \{g^{i,j}_t\}_{t \in [H]} \) for \( j \in [J] \), where \( g^{i,j}_t : \mathcal{S} \times \mathcal{A} \rightarrow [0,1] \). The threshold with respect to each player $i$'s constraints $g^{i,j}$ is denoted by $c^{i,j}\in\mathbb{R}$. We refer to the set of constraints $\{g^{i,j}\}_{j\in[J]}$ as \emph{playerwise coupling constraints} since each player has its own set of constraints that depend on the joint action profile. If, however, the constraint functions and threshold values are equal across all players, i.e., $g^{i,j}=g^j$ and $c^{i,j}=c^j$ for all $j\in[J]$ and $i\in\mc N$, then we refer to these types of constraints as \emph{common coupling constraints}, as each player is subject to the same set of constraints. For shorthand, we refer to \emph{Markov games with coupling constraints} as \emph{constrained Markov games}.

\noindent \textbf{Game dynamics } For \( t \in[H] \), let \( \pi_t : \mathcal{S} \rightarrow \Delta(\mathcal{A}) \) denote a mapping from each state \( s \in \mathcal{S} \) to a distribution over the joint action space \( \mathcal{A} \) at timestep \( t \). We define \( \pi := \{\pi_t\}_{t \in [H]} \) as a {Markovian policy}, and denote the set of all such policies by \( \Pi_M \).

In a constrained Markov game, an initial state $s_1$ is drawn from the initial distribution $\rho$, i.e., $s_1 \sim \rho\in\Delta(\mc S)$.
At each timestep $t$, the players select a joint action $a_t\sim \pi_t(\cdot\,\lvert\, s_t)$ based on policy $\pi\in\Pi_M$ and each player receives a reward $r_t^i(s_t,a_t)$ and a set of  constraints $\{g_t^{i,j}(s_t,a_t)\}_{j\in[J]}$. Then, the state $s_t$ transitions to a new state $s_{t+1} \sim P_t(\cdot\,\lvert\, s_t, a_t)$. 
The expected cumulative reward over the horizon $H$ of player $i$ is defined as:
\begin{align*}
V^{r^i}(\pi)
:= \mathbb{E}_{\substack{
s_1 \sim \rho,\, a_t \sim \pi_t(\cdot | s_t),
s_{t+1} \sim P_t(\cdot | s_t, a_t)
}}
\left[ \sum_{t=1}^{H} r^i_t(s_t, a_t) \right].
\end{align*}
and its expected cumulative constraint value  over the horizon $H$ is defined as: 
\begin{align*}
V^{g^{i,j}}(\pi)
:= \mathbb{E}_{\substack{
s_1 \sim \rho,\, a_t \sim \pi_t(\cdot | s_t),
s_{t+1} \sim P_t(\cdot | s_t, a_t)
}}
\left[ \sum_{t=1}^{H} g^{i,j}_t(s_t, a_t) \right], \,\forall j \in[J].
\end{align*}
The transition dynamics $P$ and policy $\pi\in\Pi_M$ induce a distribution over state-action pairs, the so-called state-action occupancy measure, given by: for all $t\in[H]$ and $(s_t,a_t)\in\mc S\times\mc A$,
\begin{align*}
    d_1^\pi(s_1,a_1) &= \rho(s_1)\pi_1(a_1\lvert s_1) ,\\
    d_t^\pi(s_t,a_t) &=\sum_{(s_{t-1},a_{t-1})\in\mc S\times \mc A}\!\!\!\!\!\!\!\!\!d_{t-1}^\pi(s_{t-1},a_{t-1}) P_{t-1}(s_t\lvert s_{t-1},a_{t-1})\pi_t(a_t\lvert s_t).
\end{align*}
The reward function and constraint functions can then equivalently be expressed as:
\begin{align}
    &V^{r^i}(\pi) = \sum_{t=1}^H \sum_{(s_t,a_t)\in\mc S\times \mc A} d_t^\pi(s_t,a_t) r_t^i(s_t,a_t),\label{eq:reward_constraint_via_occupancy}\\
    &V^{g^{i,j}}(\pi) = \sum_{t=1}^H \sum_{(s_t,a_t)\in\mc S\times \mc A} d_t^\pi(s_t,a_t) g_t^{i,j}(s_t,a_t), \,\forall j\in[J].\nonumber
\end{align}
In a constrained Markov game, each player $i$ is subject to constraints of the form:
\begin{align}\label{eq:i-feasible_set}
    V^{g^{i,j}}(\pi)\geq c^{i,j},\,\forall j\in[J].
\end{align}
A policy $\pi$ is said to be $i$-feasible if Inequality \eqref{eq:i-feasible_set} above holds for every $j\in[J]$. We denote the set of $i$-feasible policies by:
\begin{align}
    \mc C_\pi^i:=\{\pi\in\Pi_M\mid V^{g^{i,j}}(\pi)\geq c^{i,j},\, \forall j\in[J]\}.\label{eq:i-feasible}
\end{align}
Furthermore, a policy $\pi$ is called feasible if it is $i$-feasible for all $i\in\mc N$. We denote the set of feasible policies by:
\begin{align}
    \!\! \!\mc C_\pi:=\{\pi\in\Pi_M \mid V^{g^{i,j}}(\pi)\geq c^{i,j}, \forall j\in[J],\forall i\in\mc N\}.\label{eq:feasible_set}
\end{align}
Observe that $\mc C_\pi=\bigcap_{i\in\mc N}\mc C_\pi^i$. For Markov games with \emph{common coupling constraints}, we note that $\mc C_\pi^i$ reduces to $\mc C_\pi$ for all $i\in\mc N$ since each player has the same set of constraints.

\section{Constrained correlated equilibria}\label{sec:equinotion}

An important solution concept in Markov games is the correlated equilibrium. The constrained correlated equilibrium generalizes this notion to \emph{constrained Markov games}. Before giving a formal definition, we introduce the Markovian stochastic modification.

\begin{definition}
    A Markovian stochastic modification of player $i$ is a collection of maps $\phi^i =\{\phi_t^i\}_{t=1}^{H}$ with:
    \begin{align*}
    \phi_t^i: \mc S \times \mAi \to \Delta(\mAi),\quad\forall t\in[H].
    \end{align*}
    Denote the set of such modifications by $\Phi_M^i$. At each timestep $t$, denote $\hat a_t^i$ as player $i$'s action induced by policy $\pi_t$. Given the current state $s_t$ and player $i$'s action $\hat a_t^i$, a stochastic modification $\phi_t^i$ randomly maps $\hat a_t^i$ to another action $a_t^i$.

    A modified policy $\phi^i\circ\pi:=\{\phi_t^i\circ\pi_t\}_{t=1}^H$ is, for all $t\in[H]$, all $a_t\in\mc A$, and all $s_t\in\mc S$, defined as:
    \begin{align*}
        (\phi_t^i\circ\pi_t)(a_t\lvert s_t) =\sum_{\hat a_t^i\in\mc A^i} \phi_t^i( a_t^i\lvert s_t, \hat a_t^i)\pi_t((\hat a_t^i,a_t^{-i})\lvert s_t). 
    \end{align*}
    At each timestep $t$, an action profile $(\hat a_t^i,a_t^{-i})$ is sampled from $\pi_t$. Then $\phi_t^i$ modifies $\hat a_t^i$ to another $a_t^i$ at random.
\end{definition}
In the above definition, if $\Delta(\mc A^i)$ is replaced by $\mc A^i$, then we refer to $\phi^i=\{\phi_t^i\}_{t=1}^H$ as a \emph{Markovian deterministic modification} and denote the set of such modifications by $\Phi_{M,det}^i$. We are now ready to introduce the constrained correlated equilibrium.

\begin{definition}[Constrained correlated equilibrium] \label{def:con_CE}
    A Markovian policy $\pi\in\Pi_M$ is a constrained correlated equilibrium if $\pi$ is feasible, namely, $\pi\in\mc C_\pi$, and if for any player $i \in \mc N$ the following holds:
    \begin{align}\label{eq_notionCE}
       V^{r^i}(\pi) \ge \max_{\text{$\phi^i\in\Phi_{M}^i$ is $i$-feasible}}V^{r^i}(\phi^i \circ \pi).
    \end{align}
    In the maximization above, $\phi^i$ is said to be $i$-feasible if the modified policy $\phi^i\circ\pi $ is $i$-feasible, namely, $\phi^i\circ\pi\in\mc C_\pi^i$, where $\mc C_\pi^i$ is defined in Equation \eqref{eq:feasible_set}.
\end{definition}
In a constrained correlated equilibrium, for each player, any deviation via a stochastic modification $\phi^i_M$ is either unprofitable or leads to an infeasible policy.\looseness-1

In unconstrained normal-form games, restricting the search for profitable deviations from the set of stochastic modifications $\Phi_M^i$ to the subset of deterministic modifications $\Phi_{M,\mathrm{det}}^i$ yields an equivalent notion of correlated equilibrium. For any joint policy $\pi$, computing the best deviation for player $i$, namely \(\max_{\phi^i \in \Phi_M^i} V^{r^i}(\phi^i \circ \pi)\), is a linear program over the simplex of $\Phi_M^i$, whose extreme points correspond to $\Phi_{M,\mathrm{det}}^i$. Hence, an optimal deviation can always be chosen deterministic. In contrast, this equivalence fails in normal-form games with coupling constraints, and therefore also in Markov games with coupling constraints. In this case, the search is restricted to feasible modifications in $\Phi_M^i$, and the extreme points of the feasible set no longer correspond to $\Phi_{M,\mathrm{det}}^i$. In particular, \cite[Theorem~1]{chen2022finding} establishes that restricting to $\Phi_{M,\mathrm{det}}^i$ leads to a strictly weaker notion of constrained correlated equilibrium.

Nevertheless, we next introduce a class of modifications that preserves equivalence for constrained correlated equilibria.
\begin{theorem}\label{cor:equinotion}
The following two statements are equivalent:
\begin{enumerate}
    \item Policy $\pi$ is a constrained correlated equilibrium as per Definition~\ref{def:con_CE}.

    \item Policy $\pi$ is feasible, and for every player $i \in \mathcal N$, the following holds: for any $\alpha \in \Delta(K^i)$, where $K^i := |\Phi_{M,\mathrm{det}}^i|$, satisfying \(\sum_{k=1}^{K^i} \alpha_k V^{g^{i,j}}(\phi^i(k)\circ \pi) \geq c^{i,j}, \, \forall j \in [J]\), we have
    \[
    V^{r^i}(\pi)\geq \sum_{k=1}^{K^i} \alpha_k V^{r^i}(\phi^i(k)\circ \pi).
    \]
\end{enumerate}
We say that \( \alpha \) is $i$-\textit{feasible} if  $\alpha\in\{\alpha\in\Delta(K^i)\mid\sum_{k=1}^{K^i} \alpha_k V^{g^{i,j}}(\phi^i(k)\circ\pi) \geq c^{i,j},\,\forall j \in [J]\}$.
\end{theorem}
Theorem~\ref{cor:equinotion} shows that, in Definition~\ref{def:con_CE}, Inequality~\eqref{eq_notionCE} can be verified over convex combinations of Markovian deterministic modifications in $\Phi^i_{M,\mathrm{det}}$, rather than over all stochastic modifications in $\Phi_M^i$. In normal-form games, stochastic deviations can be restricted to deterministic ones because the objective is linear and the set is a simplex. In the presence of constraints, however, the feasible set of deviations is no longer a simplex, and optimal deviations may require randomization. Theorem~\ref{cor:equinotion} shows that it nevertheless suffices to consider mixtures over deterministic modifications. In particular, $\Phi_{M,\mathrm{det}}^i$ is finite, with cardinality \(|\Phi_{M,\mathrm{det}}^i| = H|S||\mathcal A^i|^{|\mathcal A^i|}\), whereas $\Phi_M^i$ is infinite. Hence, it suffices to optimize over a mixture $\alpha \in \Delta(K^i)$ for each player. 

The full proof of above is given in Appendix~\ref{app_proof_proposition}, and we outline a proof sketch below.

\paragraph{Proof sketch} Our goal is to reduce each player’s deviation problem to a single-agent finite-horizon MDP whose policies correspond to all possible modifications of player $i$. Once this reduction is established, we can invoke the standard result that any stochastic policy in a finite-horizon MDP is equivalent to a convex combination of deterministic policies \cite{feinberg1996measurability}.  To construct this MDP, we encode everything observable to player $i$ at time $t$ into the state, namely $(s_t, a_t^i)$, where $s_t$ is the environment state and $a_t^i$ is the recommended action. The key idea is that the \emph{action} in this MDP corresponds to a modification of the recommendation. Concretely, at state $(s_t, a_t^i)$, player $i$ selects a modified action $\hat a_t^i \sim \phi_t^i(\cdot \mid s_t, a_t^i)$. Given $(s_t, a_t^i)$ and the chosen modification $\hat a_t^i$, the remaining players act according to the original policy, i.e., $a^{-i} \sim \pi_t((\cdot,a_t^i) \mid s_t)$, and the environment transitions as $s_{t+1} \sim P_t(\cdot \mid s_t, (\hat a_t^i, a_{-i}))$. To ensure independence from $\pi$, we sample the next step recommended action $a_{t+1}^i$ uniformly from $\mc A^i$, treating it as an exogenous component of the augmented state. This defines a transition kernel $\bar P_t^\pi((s_{t+1}, a_{t+1}^i)\mid (s_t, a_t^i), \hat a_t^i)$ (see Equation~\eqref{eq_deftransition}), where the randomness over $a_t^{-i}$ is integrated out. Figure~\ref{fig:augmented-mdp} illustrates the main idea of the proof sketch. 
\begin{figure}[t]
    \centering

\begin{tikzpicture}[
    font=\sffamily\small,
    >=Latex,
    bluebox/.style={
        draw=blue!80!black,
        rounded corners=5pt,
        line width=0.9pt,
        minimum width=3.15cm,
        minimum height=0.72cm,
        align=center,
        inner sep=2pt
    },
    greenbox/.style={
        draw=green!45!black,
        rounded corners=5pt,
        line width=0.9pt,
        minimum width=4.25cm,
        minimum height=0.72cm,
        align=center,
        inner sep=2pt
    },
    bluedashed/.style={
        draw=blue!80!black,
        dashed,
        rounded corners=4pt,
        line width=0.8pt,
        minimum width=2.0cm,
        minimum height=0.58cm,
        align=center,
        inner sep=1pt
    },
    greendashed/.style={
        draw=green!45!black,
        dashed,
        rounded corners=4pt,
        line width=0.8pt,
        minimum width=4.55cm,
        minimum height=1.0cm,
        align=center,
        inner sep=2pt
    },
    bluearrow/.style={
        ->,
        line width=0.85pt,
        draw=blue!80!black,
        shorten <=2pt,
        shorten >=2pt
    },
    greenarrow/.style={
        ->,
        line width=0.85pt,
        draw=green!45!black,
        shorten <=2pt,
        shorten >=2pt
    },
    thickarrow/.style={
        -Latex,
        line width=1.4pt,
        draw=black!70
    }
]

\node[draw=none, minimum width=3.8cm, minimum height=5.4cm] (P1) at (0,0) {};
\node[draw=none, minimum width=5.4cm, minimum height=5.4cm] (P2) at (6.35,0) {};


\node[font=\bfseries\normalsize] 
    at ($(P1.north)+(0,0)$) 
    {Original game};

\node[bluebox] (s) 
    at ($(P1.north)+(0,-0.80)$) 
    {$s_t$};

\node[bluebox] (jointaction) 
    at ($(s)+(0,-1.4)$)
{
{Player $i$:} $a_t\sim \phi^i\circ\pi(\cdot\mid s_t)$
};

\node[bluedashed, minimum width=3.15cm] (transition) 
    at ($(jointaction)+(0,-1.3)$)
{
$P_t\bigl(s_{t+1}\mid s_t,a_t\bigr)$
};

\node[bluebox] (nexttime) 
    at ($(transition)+(0,-1.35)$) 
    {$s_{t+1}$};

\draw[bluearrow] (s) -- (jointaction);
\draw[bluearrow] (jointaction) -- (transition);
\draw[bluearrow] (transition) -- (nexttime);


\node[font=\bfseries\normalsize] 
    at ($(P2.north)+(0,0)$)
    {Augmented MDP for player $i$};

\node[greenbox, minimum width=2.25cm] (state) 
    at ($(P2.north)+(0,-0.80)$)
    {$(s_t,a_t^i)$};

\node[text=green!35!black] 
    at ($(state.east)+(0.50,0)$) 
    {State};

\node[greenbox, minimum width=3.35cm] (action) 
    at ($(state)+(0,-1.25)$)
    {$\hat a_t^i \sim \phi^i(\cdot\mid s_t,a_t^i)$};

\node[text=green!35!black] 
    at ($(action.east)+(0.50,0)$) 
    {Action};

\node[greendashed] (env) 
    at ($(action)+(0,-1.45)$)
{
$a_t^{-i}\sim \pi_t((\cdot, a^i)\mid s_t)$\\[-0.2mm]
$s_{t+1}\sim P_t(\cdot\mid s_t,(\hat a_t^i,a_t^{-i}))$\\[-0.2mm]
$a_{t+1}^i\sim {\small\text{Uniform distribution over }}\mc A^i$
};

\node[text=green!35!black, align=left, anchor=west] 
    at ($(env.east)+(0.18,0)$) 
    {$\bar P_t^\pi$\\[-0.5mm]\eqref{eq_deftransition}};

\node[greenbox, minimum width=2.65cm] (nextstate) 
    at ($(env)+(0,-1.50)$)
    {{\small$(s_{t+1},a_{t+1}^i)$}};

\node[text=green!35!black] 
    at ($(nextstate.east)+(0.80,0)$) 
    {Next state};

\draw[greenarrow] (state) -- (action);
\draw[greenarrow] (action) -- (env);
\draw[greenarrow] (env) -- (nextstate);


\coordinate (midL) at ($(P1.east)+(0.15,0)$);
\coordinate (midR) at ($(P2.west)+(-0.05,0)$);

\node[align=center, font=\small] (focus)
    at ($(midL)!0.5!(midR)+(0,0.30)$)
    {Focus on\\player $i$\\};

\draw[thickarrow] (midL) -- (midR);

\end{tikzpicture}

    \caption{Construction of the augmented MDP for player $i$ from the original Markov game.\looseness-1}
    \label{fig:augmented-mdp}
\end{figure}

\section{Conclusion and future work}\label{sec:conclusion}
In this paper, we studied constrained correlated equilibria in finite-horizon Markov games with coupling constraints. To this end, we first showed that constrained correlated equilibria can be characterized by convex combinations of Markovian deterministic modifications (Theorem~\ref{cor:equinotion}).

These considerations suggest several directions for future work. First, the existence result may be extended to an infinite-horizon Markov game. Another promising direction is to investigate whether common coupling constraints can avoid the computational intractability known for constrained correlated equilibria with playerwise coupling constraints~\cite{bernasconi2025complexity}. 

\bibliographystyle{splncs04}
\bibliography{ref}

\appendix
\section{Proof for Theorem \ref{cor:equinotion}}\label{app_proof_proposition}
\begin{proof}
For any player \( i \in \mathcal{N} \) and any policy \( \pi\in\Pi_M \), we construct a finite-horizon MDP defined by the tuple:
\begin{align}\label{eq:MDP2}
\left\{\bar{\mc S}\cup\{b\}, \mathcal{A}^i, H, \bar{P}^\pi, \bar{\rho}\right\}.\tag{MDP}
\end{align}
The state space is given by $\bar{\mc S}\cup\{b\}$, where $\bar{\mc S}:= \mathcal{S} \times \mathcal{A}^i$ and \( b \) is an auxiliary state. Each element of $\bar{\mc S}$ takes the form \( \bar s=(s,a^i) \). The time horizon \( H \) is the same as in the original Markov game. The action space \( \mathcal{A}^i \) corresponds to player \( i \)'s action set in the original Markov game. The transition kernel is denoted by \( \bar{P}^\pi := \{\bar{P}_t^\pi\}_{t \in [H-1]} \), where each \( \bar{P}_t^\pi : \bar{\mathcal{S}} \times \mathcal{A}^i \to \Delta(\bar{\mc S} \cup \{b\}) \) is defined as follows: 
\begin{equation}
    \begin{aligned}
    \bar{P}_t^\pi \left( (s_{t+1},a_{t+1}^i) \mid (s_t,a_t^i),\hat{a}_t^i \right)
    =& \frac{1}{|\mathcal{A}^i|} \sum_{a_t^{-i}}P_t(s_{t+1} \mid s_t, (\hat a_t^i,a_t^{-i})) \pi_t((a_t^i,a_t^{-i}) \mid s_t),\\
    \bar{P}_t^\pi \left( b \mid (s_t,a_t^i),\hat{a}_t^i \right) = &1 - \sum_{a_t^{-i}}\pi_t((a_t^i,a_t^{-i}) \mid s_t).\label{eq_deftransition}
\end{aligned}
\end{equation}
We define \( b \) to be an absorbing state, such that the transition kernel satisfies $\bar{P}_t^\pi \left( b \mid b \right) = 1$. By Lemma~\ref{lemma:MDP2_validdis} in Appendix \ref{app:proof_cor_equinotion}, we verify that $\bar P_t^\pi$ defines a valid probability distribution over $\Delta(\bar{\mathcal{S}} \cup \{b\})$ for all $t \in [H-1]$.

Moreover, we define the initial distribution for all $\bar s_1\in\bar{\mc S}\cup\{b\}$ as:
\begin{align*}
    \bar{\rho}(\bar s_1)=\begin{cases}
        \frac{\rho(s_1)}{|\mathcal{A}^i|}, \,&\text{if }\bar s_1=(s_1,a_1^i)\in\bar{\mc S},\\
        0, \,&\text{otherwise}.
    \end{cases}
\end{align*}
It is straightforward to verify that $\bar{\rho} \in \Delta(\bar{\mathcal{S}}\cup\{b\})$. Therefore, the constructed \ref{eq:MDP2} is well-posed. We further note that any Markovian stochastic modification $\phi^i\in\Phi^i_{M}$, constitutes a valid policy in the \ref{eq:MDP2}. By Lemma \ref{lemma:MDP2_eq_occ_ori} in Appendix \ref{app:proof_cor_equinotion}, for all $(s_h, (\ha_h^i,a_h^{-i})) \in \mathcal{S} \times \mathcal{A}$ and $h\in[H]$ it holds that:
\begin{align}
    d_h^{\phi^i\circ \pi}&({s}_h, (\hat{a}_h^i, a_h^{-i}))
    =|\mathcal{A}^i|^h\sum_{a_h^i}\bar d_h^{\phi^i}((s_h,a_h^i),\hat a_h^i) \pi_h((a_h^i,a_h^{-i})|s_h), \label{eq_equidistribution}
\end{align}
where $\bar d_h^{\phi^i}(\cdot)$ denotes the state-action occupancy measure at timestep $h$ of the \ref{eq:MDP2}.

Meanwhile, we leverage that Markovian stochastic policies and convex combinations of Markovian deterministic policies are equivalent in terms of the state-action occupancy measures they induce in the \ref{eq:MDP2} (see \cite[Theorem~5.2]{feinberg1996measurability}). Namely, there exists a vector $\alpha\in\Delta(K^i)$ such that:
\begin{align}\label{eq:connection_stochastic_deterministic}
    \bar d_t^{\phi^i}((s_t, a_t^i), \hat a_t^i)
    = \sum_{k=1}^{K^i} \alpha_k \, \bar d_t^{ \phi^i(k)}(( s_t, a_t^i), \hat a_t^i),
\end{align}
for all \( (s_t, a_t^i) \in \bar{\mc S} \), all \( \hat a_t^i \in \mc A^i \), and all $t\in[H]$, where each \(\phi^i(k) \in \Phi^i_{M,det} \) is a Markovian deterministic policy of the \ref{eq:MDP2}. Note that since Equation \eqref{eq:connection_stochastic_deterministic} holds with equality, the reverse statement is also true. Namely, given a vector $\alpha\in\Delta(K^i)$, there exists a $\phi^i\in\Phi_{M}^i$ such that Equation \eqref{eq:connection_stochastic_deterministic} holds.

Combining Equations \eqref{eq:connection_stochastic_deterministic} and \eqref{eq_equidistribution}, we conclude that Markovian stochastic policies and convex combinations of Markovian deterministic policies are equivalent in terms of the state-action occupancy measures they induce in the original Markov game. Since the reward $V^{r^i}$ and the constraint functions $V^{g^{i,j}}$ are linear functions of the state-action occupancy measure as seen in Equation~\eqref{eq:reward_constraint_via_occupancy}, this implies the equivalence in Theorem~\ref{cor:equinotion}.
\end{proof}
\subsection{Supporting results for Theorem \ref{cor:equinotion}}\label{app:proof_cor_equinotion}
\begin{lemma}\label{lemma:MDP2_validdis}
For $\bar P^\pi:=\{\bar P^\pi_t\}_{t\in[H-1]}$ defined in \eqref{eq:MDP2}, $\bar P^\pi_t$ is a valid probability distribution in $\Delta(\bar{\mathcal{S}}\cup \{b\})$ for all $t\in[H-1]$.  
\end{lemma}
\begin{proof}
 For any given timestep $t\in[H-1]$, state $\bar{s}_t=(s_t,a_t^i)\in\bar{S}_t$, and action $\hat{a}_t^i\in\mathcal{A}_i$, it holds that:
\begin{align*}
    &\sum_{\bar s_{t+1}\in\bar{\mc S}_{t+1}\cup\{b\}}\bar{P}_t^\pi(\bar{s}_{t+1}|\left(s_t,a_t^i\right),\hat{a}_t^i)\\
    =&\sum_{s_{t+1}}\sum_{a_{t+1}^i}\bar{P}_t^\pi ( (s_{t+1},a_{t+1}^i) \!\mid \!(s_t,a_t^i),\hat{a}_t^i )\!+\! \bar{P}_t^\pi( b \!\mid\! (s_t,a_t^i),\hat{a}_t^i )\\
    =& \frac{1}{|\mathcal{A}^i|} \sum_{s_{t+1}}\sum_{a_{t+1}^i}\sum_{a_t^{-i}}P_t(s_{t+1} \mid s_t, (\hat a_t^i,a_t^{-i})) \pi_t((a_t^i,a_t^{-i}) \mid s_t)+1 - \sum_{a_t^{-i}}\pi_t((a_t^i,a_t^{-i}) \mid s_t)\\
    =&1.
\end{align*}
Furthermore, for the auxiliary state $b$ and timestep $t\in[H-1]$, it holds that:
\begin{align*}
    &\sum_{\bar s_{t+1}\in\bar{\mc S}_{t+1}\cup\{b\}}\bar{P}_t^\pi\left(\bar{s}\mid b\right)=\bar P_t^\pi(b \mid b) = 1.
\end{align*}
We conclude that $\bar{P}_t^\pi$ is a valid probability distribution in $\Delta(\bar{\mathcal{S}}_{t+1}\cup\{b\})$.
\end{proof}

\begin{lemma}\label{lemma:MDP2_eq_occ_ori}
    For any player $i\in\mc N$, Markovian policy $\pi\in\Pi_M$, and Markovian stochastic modification \( \phi^i \in\Phi_{M}^i \), the state-action occupancy measure of the modified policy $\phi^i\circ\pi$ equals:
\begin{align}
    &d_h^{\phi^i\circ \pi}({s}_h, (\hat{a}_h^i, a_h^{-i}))
    = |\mathcal{A}^i|^h\sum_{a_h^i}\bar d_h^{\phi^i}((s_h,a_h^i),\hat a_h^i) \pi_h((a_h^i,a_h^{-i})|s_h), \label{eq_occ_ori}
\end{align}
for all $(s_h, (\ha_h^i,a_h^{-i})) \in \mathcal{S} \times \mathcal{A}$ and $h\in[H]$, where $\bar d_h^{\phi^i}(\cdot)$ denotes the state-action occupancy measure at timestep $h$ of \ref{eq:MDP2}.
\end{lemma}
\begin{proof} 
We prove the above lemma by induction.

\noindent \underline{Base case: } For $h=1$, we have that for any $(s_1,(\hat{a}_1^i,a_1^{-i}))\in \mc S\times\mc A$ the following holds:
\begin{align*}
    &|\mathcal{A}^i| \sum_{a_1^{i}} \bar{d}_1^{\phi^i}((s_1,a_1^i),\hat a_1^i)\pi_1((a_1^i,a_1^{-i}) \mid s_1)\\
    \overset{(i)}{=}& |\mathcal{A}^i| \sum_{a_1^{i}}\bar{\rho}(s_1,a_1^i) \phi^i_1(\ha^i_1|s_1,a_1^i)\pi_1((a_1^i,a_1^{-i}) \mid s_1)\\
    \overset{(ii)}{=}&\sum_{a_1^{i}} \rho(s_1)\phi_1^i(\ha^i_1|s_1,a_1^i )\pi_1((a_1^i,a_1^{-i}) \mid s_1)\\
    =& d_1^{\phi^i\circ \pi}({s}_1, (a_1^{-i},\hat a_1^i)),
\end{align*}
where in step $(i)$ we express \( \bar{d}_1^{\phi^i} \) by the dynamics of \ref{eq:MDP2}. In step \((ii)\) we plug in the definition of \(\bar{\rho}\). 
\\
\\
\noindent \underline{Induction step: } Assume that Equation \eqref{eq_occ_ori} holds for all \( (s_h, (a_h^{-i},\ha_h^i)) \in \mathcal{S} \times \mathcal{A} \) and all \( h \in [t-1] \), then we show that Equation \eqref{eq_occ_ori} also holds for $t$. We can rewrite \(\bar{d}_t^{\phi^i}\) as:
\begin{align*}
    &\bar{d}_t^{\phi^i}((s_t,a_t^i),\hat a_t^i)\\
    \overset{(i)}{=}&\phi^i_t(\ha_t^i|s_t,a_t^i)\sum_{ \bar s_{t-1}}\sum_{\bar a_{t-1}^i}\sum_{\hat a_{t-1}^i}\bar{d}_{t-1}^{\phi^i}((\bar s_{t-1},\bar a_{t-1}^i),\hat a_{t-1}^i)\bar{P}_{t-1}^\pi(s_t,a_t^i|(\bar s_{t-1},\bar a_{t-1}^i),\hat{a}_{t-1}^i)\\
    \overset{(ii)}{=}&|\mathcal{A}^i|^{-1}\phi^i_t(\ha_t^i|s_t,a_t^i)\sum_{ \bar s_{t-1}}\sum_{\bar a_{t-1}^i}\sum_{\hat a_{t-1}^i}\sum_{\bar a_{t-1}^{-i}}\bar{d}_{t-1}^{\phi^i}((\bar s_{t-1},\bar a_{t-1}^i),\hat a_{t-1}^i)\\
    &\times \pi_{t-1}((\bar a_{t-1}^i,\bar a_{t-1}^{-i}) \mid \bar s_{t-1}) P_{t-1}(s_t|\bar s_{t-1},(\ha_{t-1}^i,\bar a_{t-1}^{-i}))\\
    \overset{(iii)}{=}&|\mathcal{A}^i|^{-t}\phi^i_t(\ha_t^i|s_t,a_t^i)\sum_{ \bar s_{t-1}}\sum_{\hat a_{t-1}^i}\sum_{\bar a_{t-1}^{-i}}d_{t-1}^{\phi^i\circ\pi}(\bar s_{t-1},(\hat a_{t-1}^i,\bar a_{t-1}^{-i})) \\
    & \times P_{t-1}(s_t|\bar s_{t-1},(\ha_{t-1}^i,\bar a_{t-1}^{-i})).
\end{align*}
In step \((i)\), we express the state-action distribution using the transition dynamics, in step \((ii)\), we apply the definition of \(\bar{P}^\pi_{t-1}\), and in step \((iii)\), we apply Equation \eqref{eq_occ_ori} which holds by induction for timestep $t-1$. Using the above equation, we obtain:
\begin{align*}
    &|\mathcal{A}^i|^t\sum_{a_t^i}\bar{d}_t^{\phi^i}((s_t,a_t^i),\hat a_t^i) \pi_t((a_t^i,a_t^{-i})|s_t)\\
    =&\sum_{a_t^i}\phi^i_t(\ha_t^i|s_t,a_t^i)\sum_{ \bar s_{t-1}}\sum_{\hat a_{t-1}^i}\sum_{\bar a_{t-1}^{-i}}d_{t-1}^{\phi^i\circ\pi}(\bar s_{t-1},(\hat a_{t-1}^i,\bar a_{t-1}^{-i})) P_{t-1}(s_t|\bar s_{t-1},(\ha_{t-1}^i,\bar a_{t-1}^{-i}))\\
    & \times \pi_t((a_t^i,a_t^{-i})|s_t)\\
    =&d_t^{\phi^i\circ\pi}(s_t,(\hat a_t^i,a_t^{-i})),
\end{align*}
where the last step is followed by the definition of state-action occupancy measure.
\end{proof}

\end{document}